\documentclass[aps,prapplied,reprint,groupedaddress]{revtex4-2}

\usepackage{graphicx}
\usepackage{dcolumn}
\usepackage{bm}
\usepackage[utf8]{inputenc}
\usepackage[T1]{fontenc}
\usepackage{mathptmx}
\usepackage{etoolbox}
\usepackage{siunitx,amsmath,amssymb}
\usepackage{braket}
\usepackage[shortlabels]{enumitem}
\usepackage[unicode,pdfusetitle,colorlinks,allcolors=blue]{hyperref}
\hypersetup{pdfauthor={M. Mariantoni}}

\begin{document}

\title[Quantum Socket and DemuXYZ-Based Gates]{The Quantum Socket and DemuXYZ-Based Gates with Superconducting Qubits}

\author{J. H.~B\'{e}janin}
\affiliation{Institute for Quantum Computing, University of Waterloo, 200 University Avenue West, Waterloo, Ontario N2L 3G1, Canada}
\affiliation{Department of Physics and Astronomy, University of Waterloo, 200 University Avenue West, Waterloo, Ontario N2L 3G1, Canada}

\author{C. T.~Earnest}
\affiliation{Institute for Quantum Computing, University of Waterloo, 200 University Avenue West, Waterloo, Ontario N2L 3G1, Canada}
\affiliation{Department of Physics and Astronomy, University of Waterloo, 200 University Avenue West, Waterloo, Ontario N2L 3G1, Canada}

\author{M.~Mariantoni}
\email[Corresponding author: ]{matteo.mariantoni@uwaterloo.ca}
\affiliation{Institute for Quantum Computing, University of Waterloo, 200 University Avenue West, Waterloo, Ontario N2L 3G1, Canada}
\affiliation{Department of Physics and Astronomy, University of Waterloo, 200 University Avenue West, Waterloo, Ontario N2L 3G1, Canada}

\date{\today}

\begin{abstract}
Building large-scale superconducting quantum computers requires two complimentary elements: scalable wiring techniques and multiplex architectures. In our previous work [Béjanin \textit{et al.}, \href{https://doi.org/10.1103/PhysRevApplied.6.044010}{Phys. Rev. Applied \textbf{6}, 044010 (2016)}], we have introduced and characterized a truly vertical interconnect named the \emph{quantum socket}. In this paper, we exercise the quantum socket using high-coherence flux-tunable Xmon transmon qubits. In particular, we test potential qubit heating and one-qubit gate performance. We observe no heating effects and time-stable gate fidelities in excess of~\SI{99.9}{\percent}. We then propose and experimentally characterize a demultiplexed gate technique based on flux pulses and a common continuous drive signal: \emph{DemuXYZ}. We discuss DemuXYZ's working principle, show its operation, and perform quantum process tomography on a selection of one-qubit gates to confirm proper operation. We obtain fidelities around~\SI{93}{\percent} likely limited by flux-pulse imperfections. We finally discuss future solutions for wiring integration as well as improvements to the DemuXYZ technique.
\end{abstract}

\maketitle

\section{\label{sec:Introduction}Introduction}

Superconducting quantum computers have reached medium-scale integration, with up to about a hundred qubits. These computers have been used to demonstrate quantum advantage~\cite{Arute:2019} and to take the first steps toward experimental quantum error correction~\cite{Krinner:2022,Acharya:2022} as well as quantum simulations~\cite{Karamlou:2022,Andersen:2022}. Despite these successes, building large-scale quantum computers with thousands or millions of qubits, while maintaining low error rates, requires truly \emph{scalable wiring} techniques as well as control and readout \emph{multiplexing}.

State-of-the-art flip-chip techniques have made it possible to build medium-scale quantum computers by allowing access to the center regions of two-dimensional qubit arrays~\cite{Rosenberg:2017,Foxen:2017,Kosen:2022}; however, these techniques still rely on lateral wire bonds at the four edges of the flipped chip, hindering further scalability. In fact, in a two-dimensional~$N \times N$ array, the number of qubits scales quadratically with~$N$, but the real estate available to wire bonds only scales proportionally to~$4N$. In our previous work of Ref.~\cite{Bejanin:2016}, we have demonstrated a truly-vertical interconnect based on three-dimensional wires--the quantum socket--which links the external cables required to control and readout qubits to any location on a qubit chip. This interconnect does not suffer from the limitations inherent to lateral wire bonds, as it makes it possible to directly access all~$N^2$ qubits in the array.

In this paper, we use the quantum socket to perform experiments on frequency-tunable Xmon transmon qubits~\cite{Barends:2013}. We first test the near-dc performance of the three-dimensional wires by tuning a qubit over a wide frequency bandwidth. These experiments allow us to verify whether any wire or contact resistance impedes the operation of the quantum socket due to heat generation. Our results show no detectable heating effects on qubits. We then perform randomized benchmarking~(RB) experiments to test one-qubit gate performance. We measure fidelities in excess of~\SI{99.9}{\percent}, with the coherent portion of the error accounting for half of the total error. In addition, we monitor qubit stochastic fluctuations with continuous~RB experiments over a time period of~\SI{40}{\hour}, showing a stable operation with fidelities always higher than~\SI{99.9}{\percent}. Together with the detailed characterization reported in our previous work of Ref.~\cite{Bejanin:2016}, the results presented here indicate that the quantum socket is a viable interconnect for superconducting quantum computing architectures. Notably, the usage of the quantum socket is not confined to the realm of superconducting quantum computing, as the socket can also be employed by practitioners working, for example, on semiconductor-based systems.

The development of a truly scalable wiring technique is necessary to build a large-scale quantum computer; however, achieving ever larger number of qubits also requires a way to multiplex the control and readout signals carried by those wires. When implemented, quantum multiplexing would be analogous to its classical counterpart, where billions of transistors are controlled by a mere couple of thousands of external wires.

In this paper, we first propose a technique to realize one- and two-qubit gates for frequency tunable qubits--DemuXYZ--where instead of~$2N$ wires to control~$N$ qubits, only $N+1$ wires are needed. DemuXYZ is based on flux pulses that bring a qubit on resonance with a common continuous drive signal. This signal makes it possible to drive qubit transitions, while the time duration and length of the flux pulses are used to control the phase. We then experimentally implement one-qubit DemuXYZ-based gates by demonstrating the necessary tuneup procedure. We finally characterize the fidelity of such gates by means of quantum process tomography~(QPT). We find fidelities as high as~\SI{96}{\percent}, which can be likely improved by suitable pulse engineering. As for the quantum socket, the working principle of the DemuXYZ technique can be transferred to quantum computing implementations other than superconducting qubits.

It is worth noting that, besides the quantum socket, only one other truly vertical interconnect has been proposed and characterized, as shown in the works of Refs.~\cite{Rahamim:2017,Spring:2022}. While spatial demultiplexing has been proposed in the work of Ref.~\cite{Versluis:2017}, quantum multiplexing is still in its embryonic stage. It is therefore important to further develop both wiring and multiplexing to build practical quantum computers.

This paper is organized as follows. In Sec.~\ref{sec:The:Quantum:Socket}, we test for potential heating effects and assess gate performance of the quantum socket. In Sec.~\ref{sec:DemuXYZ-Based:Gates}, we introduce the DemuXYZ technique and explain its working principles. We additionally confirm the proper operation of DemuXYZ-based gates by means of~QPT. In Sec.~\ref{sec:Discussion}, we discuss current and future perspectives related to scalable wiring techniques and multiplex architectures. Finally, in Sec.~\ref{sec:Conclusions}, we summarize our results.

\section{\label{sec:The:Quantum:Socket}The Quantum Socket}

The quantum socket is a wiring technique based on spring-loaded coaxial pins. The inner and outer diameters of the pins are approximately~$400$ and \SI{1300}{\micro\meter}, respectively. The pins are designed to have a characteristic impedance of~\SI{50}{\ohm}, with a measured impedance mismatch of less than~\SI{5}{\ohm}. Such a mismatch results in a reflection coefficient of approximately~\SI{-15}{dBm} around~\SI{6}{\giga\hertz}. The pins mate with an on-chip contact pad, providing electrical connection from dc to more than~\SI{10}{\giga\hertz}. The quantum socket is made from nonmagnetic materials and works at temperatures below~\SI{10}{\milli\kelvin}. A detailed description of the quantum socket can be found in our previous work of Ref.~\cite{Bejanin:2016}. It is worth mentioning that we have been using the quantum socket in resonator and qubit experiments since~$2016$, without experiencing any reliability problems.

In Sec.~\ref{subsec:Heating:characterization:with:flux-tuning:signals}, we characterize possible heating effects in the quantum socket and in Sec.~\ref{subsec:Randomized:benchmarking:and:temporal:stability} its performance with one-qubit~RB, including temporal stability.

\subsection{\label{subsec:Heating:characterization:with:flux-tuning:signals}Heating characterization with flux-tuning signals}

\begin{figure*}[!ht]
	\centering
	\includegraphics{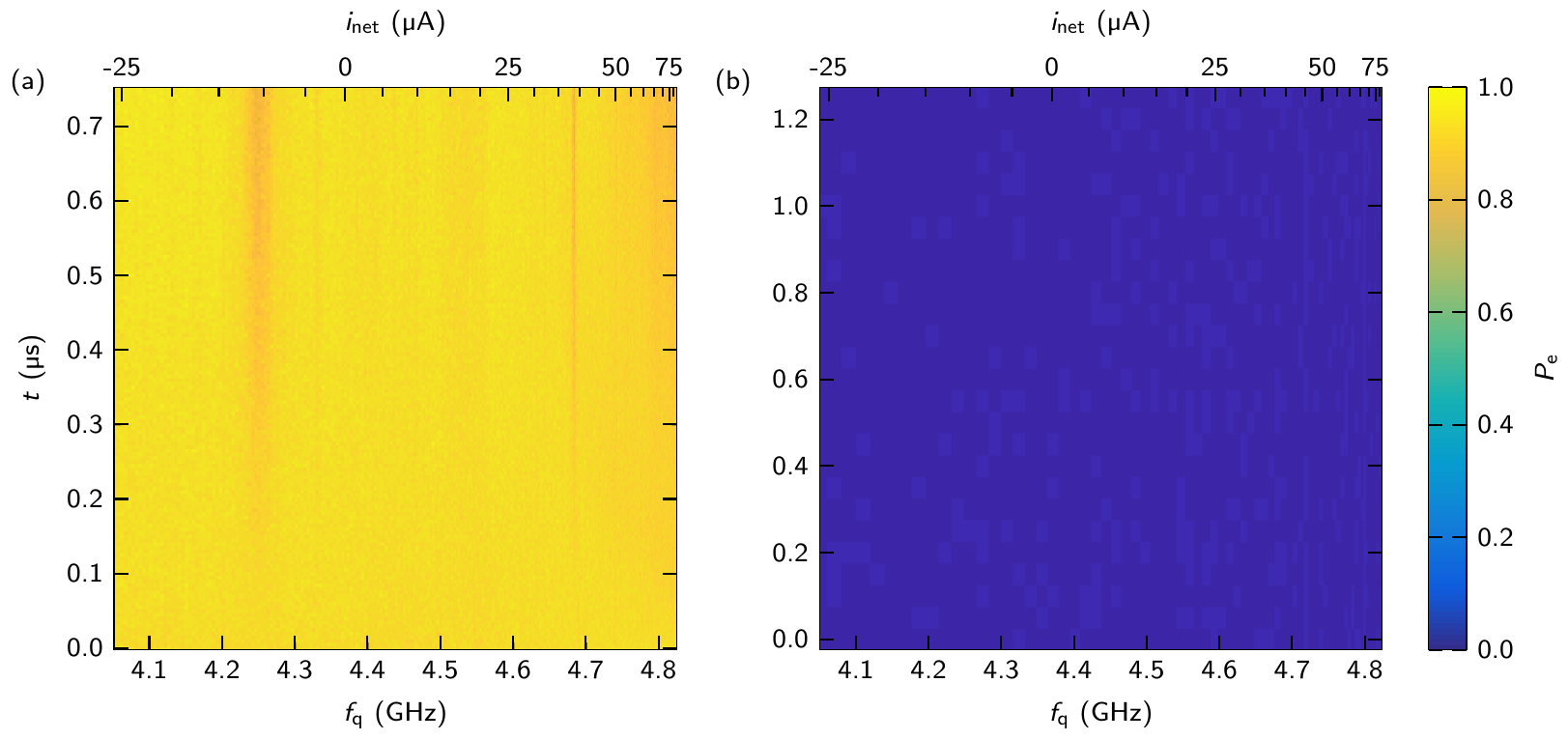}
	\caption{Swap spectroscopy experiments. $P_{\text{e}}$~(color bar)~vs.~$f_{\text{q}}$ and $t$, where~$t$ is the flux-pulse duration. We show the dependence on~$i_{\text{net}}$ on the top axes. (a)~Qubit prepared in the excited state~$\ket{\text{e}}$ (standard swap spectroscopy). (b)~Qubit prepared in the ground state~$\ket{\text{g}}$ (ground-state swap spectroscopy). We note that the two experiments are conducted with different frequency and time resolutions.}
	\label{Figure01-Bejanin_2022b_Sub}
\end{figure*}

\begin{figure}[!hb]
	\centering
	\includegraphics[width=\columnwidth]{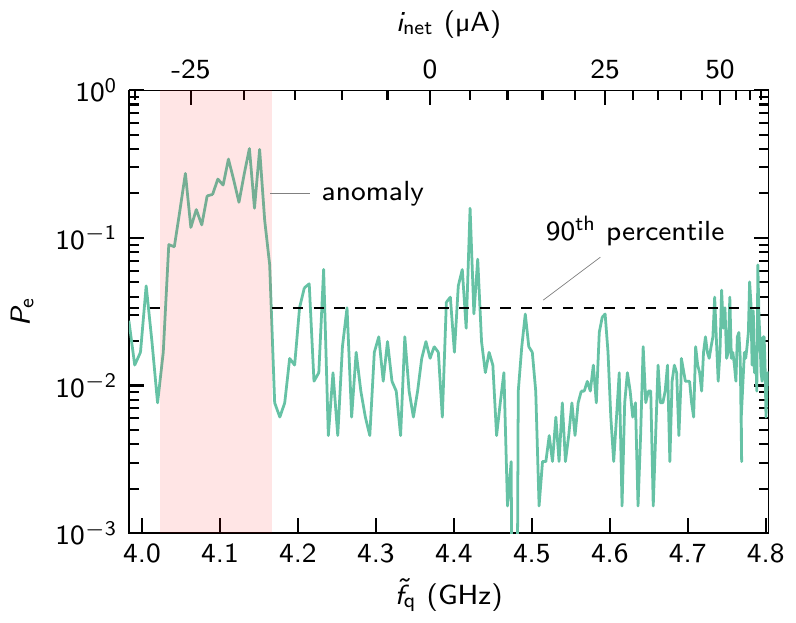}
	\caption{Idle-frequency characterization. $P_{\text{e}}$~vs.~$\tilde{f}_{\text{q}}$ with the qubit prepared in~$\ket{\text{g}}$. We show the dependence on~$i_{\text{net}}$ on the top axis. The horizontal dashed black line indicates the~$90^{\text{th}}$ percentile, which is computed by excluding the shaded band.}
	\label{Figure02-Bejanin_2022b_Sub}
\end{figure}

In this work, we perform experiments using frequency-tunable Xmon transmon qubits. The main qubit parameters, such as the qubit energy relaxation time~$T_1$, are reported in Sec.~S1 within the Supplemental Material~\cite{SM}. The qubit transition frequency~$f_{\text{q}}$ can be tuned by applying an external flux bias to the qubit SQUID loop. This flux is generated by a current ideally going through a perfect galvanic connection between the quantum socket pins and the on-chip pads; however, in presence of any pin-pad contact resistance, this current may result in unwanted chip heating. Heating, in turn, leads to thermal qubit population and, therefore, qubit initialization errors.

We characterize possible heating effects in the quantum socket by performing standard and ground-state swap spectroscopy experiments. In standard swap spectroscopy~\cite{Mariantoni:2011}, $T_1$ is measured over a range of~$f_{\text{q}}$ values. Ground-state swap spectroscopy follows a similar protocol, except that the qubit is never energized with a~$\pi$-pulse and, thus, should ideally remain in the ground state at all times. These experiments allow us to measure the excited-state population~$P_{\text{e}}$ as a function of time~$t$ and for different flux bias amplitudes; this population is labeled~$P_{\text{e} | \text{e}}$ or $P_{\text{e} | \text{g}}$ depending on whether the qubit is prepared in the excited or ground state. In the absence of any significant heating, $P_{\text{e} | \text{e}}$ and $P_{\text{e} | \text{g}}$ should remain close to one.

In order to implement the swap spectroscopy experiments, we first calibrate the qubit~$\pi$-pulse and readout at the qubit idle frequency~$\tilde{f}_{\text{q}}$. This frequency is a specific value of~$f_{\text{q}}$ chosen such that the qubit has the best control and readout characteristics; $\tilde{f}_{\text{q}}$ is exclusively controlled with a quasistatic voltage bias generated by a battery, corresponding to a current~$i_{\text{idle}}$ going through the pin-pad connection. We then sweep~$f_{\text{q}}$ by applying a rectangular voltage pulse, corresponding to a current~$\delta i$. This pulse is generated by an arbitrary waveform generator~(AWG) and is linearly superposed with the quasistatic bias by means of a custom bias tee. For every value of~$f_{\text{q}}$, the net voltage corresponds to a particular value of current~$i_{\text{net}}=i_{\text{idle}}+\delta i$, going through the pin-pad connection. The repetition rate of each pulse sequence is~\SI{2}{\kilo\hertz}, resulting in a flux-pulse duty cycle of~\SI{0.2}{\percent} for a pulse duration of~\SI{1}{\micro\second} (see Fig.~\ref{Figure01-Bejanin_2022b_Sub} for other pulse-duration values). This type of pulse sequence exemplifies the majority of flux-pulse protocols used in quantum computations, for example, flux-tuned two-qubit gates.

In certain experiments, it may be necessary to quasistatically bias the qubit at high values of~$i_{\text{net}}$ (i.e., a duty cycle of~\SI{100}{\percent}). In order to characterize this scenario, we perform an experiment where we measure~$P_{\text{e} | \text{g}}$ over a range of idle biases. Unlike swap spectroscopy, this experiment requires calibrating the qubit~$\pi$-pulse and readout at each bias point.

Figure~\ref{Figure01-Bejanin_2022b_Sub} presents the results for the standard and ground-state swap spectroscopy experiments. Figure~\ref{Figure01-Bejanin_2022b_Sub}~(a) shows that~$\langle P_{\text{e} | \text{e}}(f_{\text{q}},t=0) \rangle \approx 0.92(1)$ and Fig.~\ref{Figure01-Bejanin_2022b_Sub}~(b) that~$\langle P_{\text{e} | \text{g}}(f_{\text{q}},t=0) \rangle \approx 0.990(4)$ averaged over all values of~$f_{\text{q}}$. These findings indicate good qubit initialization. In both experiments, $P_{\text{e} | \text{e}}$ and $P_{\text{e} | \text{g}}$ remain close to their initial values at all times. Notably, $P_{\text{e} | \text{e}}$ is measured only up to~\SI{750}{\nano\second} and, thus, the~$T_1$ time decay due to qubit energy relaxation is barely visible.

In these experiments, sweeping~$f_{\text{q}}$ from approximately~$4.051$ to~\SI{4.825}{\giga\hertz} is achieved by applying a current between~$i_{\text{net}} \approx -25.8$ and~\SI{88.3}{\micro\ampere}; it is worth noting that, due to the cosine term in the transmon qubit Hamiltonian, the relationship between~$f_{\text{q}}$ and $i_{\text{net}}$ is nonlinear (see top and bottom axes in figure). Despite a heavy magnetic shielding and a meticulous choice of nonmagnetic components in the setup, our experiments are affected by a constant flux offset; thus, when the external flux is zero, instead of finding~$f_{\text{q}}$ at its maximum value, we measure~$f_{\text{q}} \approx \SI{4.369}{\giga\hertz}$. The qubit idle frequency is~$\tilde{f}_{\text{q}} \approx \SI{4.644}{\giga\hertz}$, where~$i_{\text{net}} \approx \SI{32}{\micro\ampere}$.

Joule heating would result in a gradual decrease of~$P_{\text{e} | \text{e}}$ due to lower values of~$T_1$ at higher temperatures. This effect should become noticeable for a long flux-pulse duration and high values of~$i_{\text{net}}$. In Fig.~\ref{Figure01-Bejanin_2022b_Sub}~(a), we should thus observe two triangular regions with lower values of~$P_{\text{e} | \text{e}}$ in the top-left and top-right corners of the heat map. While the top-left corner does not indicate any heating, we notice a slight triangular shade in the top-right corner; however, this is due to the Purcell effect associated with the readout resonator at~$f_{\text{m}} \approx \SI{5.032}{\giga\hertz}$. Additionally, we observe a couple of stripes with lower~$P_{\text{e} | \text{e}}$ in the middle of the heat map (at approximately~$4.250$ and \SI{4.683}{\giga\hertz}); these features are due to two-level systems~(TLSs) interacting with the qubit.

The ground-state swap spectroscopy is better suited than standard swap spectroscopy to characterize Joule heating, as it is less susceptible to decoherence phenomena due to TLSs or Purcell effect. For this experiment, heating would result in an increase of~$P_{\text{e} | \text{g}}$ for a long flux-pulse duration and high values of~$i_{\text{net}}$. We do not observe any noticeable change in~$P_{\text{e} | \text{g}}$ over the entire heat map.

We further characterize possible heating effects that could result from the quasistatic bias alone and display the results in Fig.~\ref{Figure02-Bejanin_2022b_Sub}. Unlike swap spectroscopy, this experiment is susceptible to frequency-dependent performance variations of the qubit readout; these variations are unrelated to the flux bias. For example, around~$f_{\text{q}} \approx \SI{4.1}{\giga\hertz}$ we observe an anomalous increase in~$P_{\text{e}}$; however, this effect is not due to heating since~$P_{\text{e}}$ decreases below approximately~\SI{4.02}{\giga\hertz}, which is obtained at a higher bias current. Importantly, \SI{90}{\percent} of all data points shown in this experiment, excluding the anomalous regions highlighted by the shaded band in Fig.~\ref{Figure02-Bejanin_2022b_Sub}, have~$P_{\text{e} | \text{g}} < 0.034$. This indicates a very good initialization in~$\ket{\text{g}}$ and, thus, negligible heating errors. Additionally, we note that good initialization implies the qubit chip is well thermalized to the base temperature of our dilution refrigerator~(DR) by means of the (normal conducting) three-dimensional wires of the quantum socket; the estimated qubit temperature is approximately~\SI{70}{\milli\kelvin}.

\subsection{\label{subsec:Randomized:benchmarking:and:temporal:stability}Randomized benchmarking and temporal stability}

\begin{figure*}[!ht]
	\centering
	\includegraphics{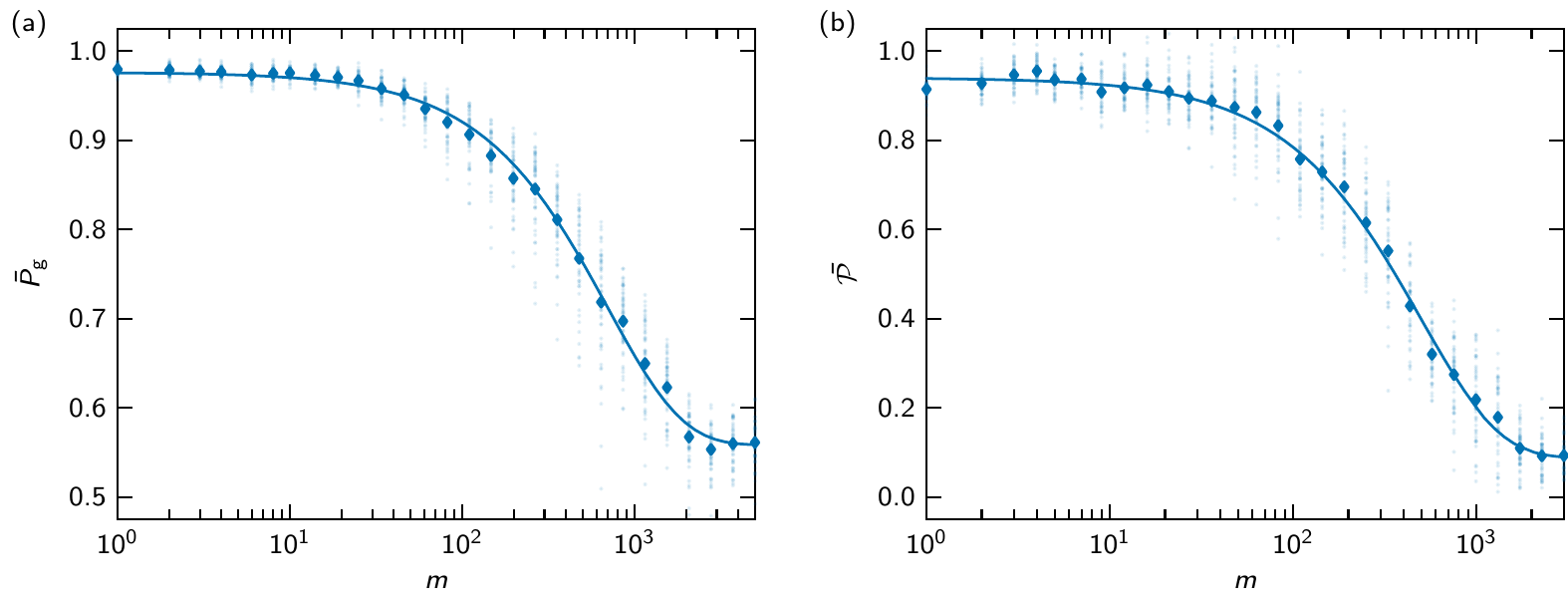}
	\caption{RB experiments. (a)~Standard~RB experiment. $\bar{P}_{\text{g}}$ vs.~$m$ (diamonds). The semitransparent dots represent~$P_{\text{g}}$ values from individual random sequences ($50$ for each~$m_i$). The solid line is the fitting curve to the model of Eq.~(\ref{eq:Pgmbar}), with fitting parameters~$p=0.99857(5)$, $A=0.418(5)$, and $B=0.558(5)$. (b)~PB experiment. $\bar{\mathcal{P}}$ vs.~$m$ (diamonds). The semitransparent dots represent~$\mathcal{P}$ values from individual random sequences ($50$ for each~$m_i$). The solid line is the fitting curve to the model of Eq.~(\ref{eq:PPmbar}), with fitting parameters~$u=0.99798(9)$, $A^{\prime}=0.85(1)$, and $B^{\prime}=0.09(1)$. The standard errors of the fit are indicated in parentheses.}
	\label{Figure03-Bejanin_2022b_Sub}
\end{figure*}

We characterize the quantum socket performance with one-qubit~RB, which allows us to estimate the average gate fidelity when applying control pulses through the three-dimensional wires. The standard~RB protocol~\cite{Emerson:2005,Dankert:2009,Magesan:2011,Boone:2019} makes use of progressively longer sequences of random gates~$G_1,G_2,...,G_m$ designed to twirl the state of one qubit (or a system of multiple qubits). At the end of each sequence, we apply a final recovery gate~$G_{m+1}$ that inverts the sequence, thereby returning the qubit to~$\ket{\text{g}}$. Given a perfect qubit and perfect gates, the RB~protocol would always result in a readout of~$\ket{\text{g}}$ for any value of~$m$. In presence of errors (e.g., qubit energy relaxation and dephasing or unitary gate errors), the protocol is equivalent to a simple depolarizing channel. As a result, the qubit is not always returned to~$\ket{\text{g}}$ and, on average, $P_{\text{g}}=1-P_{\text{e}}$ decays exponentially with~$m$. We define the quantity~$\bar{P}_{\text{g}}(m)$ as the average of~$P_{\text{g}}$ over many random sequences of length~$m$ and fit to the model
\begin{equation}
	\bar{P}_{\text{g}}(m) = A p^m + B ,
	\label{eq:Pgmbar}
\end{equation}
where~$p$, $A$, and $B$ are fit parameters. These parameters lead to an estimate of the average gate fidelity
\begin{equation}
	\bar{F} = \dfrac{1}{d} + p \left( 1 - \dfrac{1}{d} \right) ,
\end{equation}
where~$d=2$ for a single qubit.

The random gates are drawn from the Clifford group, which allows the recovery gate to be efficiently calculated, even for many qubits. The one-qubit Clifford group comprises~$24$ gates that can be defined as rotations about an axis of the Bloch sphere. In order to experimentally implement these gates, each Clifford must be decomposed in terms of \emph{primitive} gates that can be physically executed. In the case of superconducting qubits, there are two types of physical gates: $X$ (or $Y$) gates based on microwave pulses and $Z$ gates based on flux pulses; however, $Z$ gates can be realized in software as virtual gates.

Here, we use a primitive set based on physical~$X$ and virtual~$Z$ gates. Note that there might be multiple decompositions for a given Clifford. If this occurs when generating RB~sequences, we pick a decomposition at random. On average, this decomposition method results in~$23/24 \approx 0.9583$ microwave pulses per Clifford. More details on this method, including exemplary decompositions for each one-qubit Clifford gate, are found in App.~\ref{app:Clifford:Gate:Decomposition}.

Several parameters must be chosen to run an~RB experiment: The list of progressively longer sequence lengths~$L_{\text{S}} = \{m_1, m_2,...,m_i,...,m_M\}$, where~$M$ is the number of elements in the list, and the number~$N$ of random sequences to run for each length~$m_i$. In our experiments, we set~$N=50$, which is sufficient to obtain a good fit from Eq.~(\ref{eq:Pgmbar}). The sequence lengths~$m_i \in L_{\text{S}}$ are to be picked according to the physical system being benchmarked. For example, a high-performance qubit might require very long sequences to display a clear exponential decay in~$\bar{P}_{\text{g}}(m)$.

\begin{figure*}[!ht]
	\centering
	\includegraphics{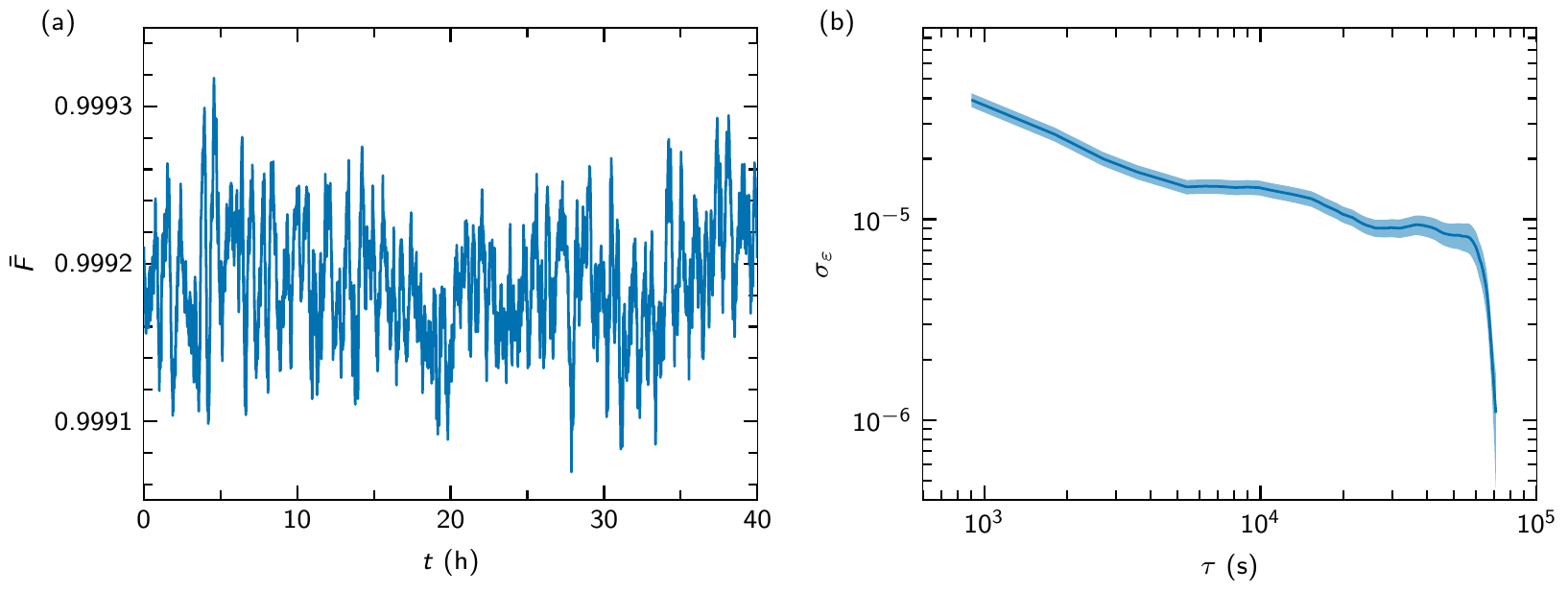}
	\caption{RB temporal stability. (a)~$\bar{F}$ vs.~$t$. The standard deviation of~$\bar{F}$, approximately~$\num{3.8e-5}$, is larger than the average fitting error, which is~\num{2.5(7)e-5}. This confirms that the fluctuations are not a fitting artifact. (b)~AD of the gate infidelity, $\sigma_{\varepsilon}$, vs.~$\tau$. The shaded area indicates the standard deviation error of the~AD.}
	\label{Figure04-Bejanin_2022b_Sub}
\end{figure*}

Given the above parameters, the experimental protocol is as follows: For each length~$m_i \in L_{\text{S}}$,
\begin{enumerate}[(I)]
	\item Draw~$m_i$ Clifford gates at random (for a single qubit, this amounts to drawing random integers between~$1$ and $24$).
	\item Compute the recovery Clifford gate.
	\item Decompose the full sequence, including the recovery gate, into a sequence of primitive gates.
	\item Execute the primitive gate sequence and read out the qubit (this step is repeated many times to get a good estimate for~$P_{\text{g}}$).
	\item Repeat steps~(I) to (IV) $N$ times in order to calculate the average~$\bar{P}_{\text{g}}(m_i)$.
\end{enumerate}

RB is easy to implement experimentally; however, it only results in a single performance metric about a qubit system, that is, the total error~$\epsilon=1-\bar{F}$. Several extensions have therefore been proposed to obtain a more thorough system characterization. One such extension is purity benchmarking~(PB), which gives an estimate of the incoherent error~$\epsilon_{\text{inc}}$~\cite{Wallman:2015,Feng:2016}. Executing~RB and PB allows us to estimate both coherent and incoherent errors and, thus, to pinpoint whether the gate performance is limited by decoherence or physical implementation.

A~PB experiment is similar to the standard~RB experiment, with the difference that instead of measuring only~$P_{\text{g}}$, which is equivalent to~$\left(1 + \braket{\hat{\sigma}_z}\right)/2$, we also measure~$\braket{\hat{\sigma}_x}$ and $\braket{\hat{\sigma}_y}$. It is thus necessary to run each sequence two additional times with an extra final~$\pi/2$ $Y$ or $X$ gate. Measuring these three expectation values makes it possible to estimate the \emph{purity} of the final state for each sequence,
\begin{equation}
	\mathcal{P} = \braket{\hat{\sigma}_z}^2 + \braket{\hat{\sigma}_y}^2 + \braket{\hat{\sigma}_z}^2 .
\end{equation}

The average value of~$\mathcal{P}$ over all sequences is also modeled to decay exponentially with~$m$ as
\begin{equation}
	\bar{\mathcal{P}}(m) = A^{\prime} u^{m-1} + B^{\prime} ,
	\label{eq:PPmbar}
\end{equation}
where~$u$, $A^{\prime}$, and $B^{\prime}$ are fit parameters. These parameters lead to an estimate for the incoherent error
\begin{equation}
	\epsilon_{\text{inc}} = \frac{1}{2} \left(1 + \sqrt{u}\right) .
\end{equation}

Given~$\epsilon$ and $\epsilon_{\text{inc}}$, we can calculate the coherent error
\begin{equation}
	\epsilon_{\text{coh}} = \epsilon - \epsilon_{\text{inc}} .
\end{equation}
Note that, since we measure~$\bar{P}_{\text{g}}$ during a~PB experiment, we can directly estimate~$\epsilon$ from the same data set.

Superconducting qubits are affected by stochastic fluctuations in~$T_1$ and $f_{\text{q}}$ due to the interaction with TLSs~\cite{Bejanin:2021:a}. Consequently, a single realization of experiments such as~RB or PB is an unreliable predictor of the qubit's performance range. It has therefore become customary to present data from experiments repeated over long time periods, rather than a single data point from a so-called ``hero'' run.

We elect to characterize fluctuations with standard~RB rather than~PB since the latter decreases the experimental repetition rate by a factor of three. In order to measure~$\bar{F}$ as a function of time, we slightly modify the protocol presented above. Instead of measuring~$P_{\text{g}}$ for~$N=50$ random sequences for each~$m_i$, we measure it for~$N=1$. When all~$M$ lengths are measured, which takes a fixed time~$\Delta t$, we start the loop over again and repeat for as long as desired; we record the start time~$t_j$ of each iteration. We then calculate~$\bar{F}$ at each~$t_j$ by computing the average~$\bar{P}_{\text{g}}(m_i)$ over a moving window of~$n$ iterations centered on~$t_j$.

Figures~\ref{Figure03-Bejanin_2022b_Sub} and \ref{Figure04-Bejanin_2022b_Sub} present the results for the~RB experiments (see Sec.~S1 within the Supplemental Material for details about qubit~$T_1$ and qubit dephasing~$T_2$~\cite{SM}). Figure~\ref{Figure03-Bejanin_2022b_Sub}~(a) shows the results of a standard~RB experiment. We choose~$M=28$ sequence lengths logarithmically spaced between~$1$ and $5000$ Clifford gates and average~$\bar{P}_{\text{g}}(m_i)$ over~$N=50$. Microwave pulses have a fixed length of~\SI{20}{\ns}, with a full width at half maximum of~\SI{13}{\nano\second}, and a DRAG-optimized Gaussian shape~\cite{Motzoi:2009,Gambetta:2011}. Notably, we do not include any buffer time between pulses. The longest sequences have a time length of approximately~$5000 \times \SI{20}{\nano\second} = \SI{100}{\micro\second}$~\footnote{This time length is approximate because the number of gates is not generally equal to the number of pulses.}. We fit the~$\bar{P}_{\text{g}}$ decay to Eq.~(\ref{eq:Pgmbar}) and obtain~$\bar{F} = \SI{99.929(3)}{\percent}$, or, equivalently, $\epsilon = 0.00071(3)$.

\begin{figure*}[!ht]
	\centering
	\includegraphics{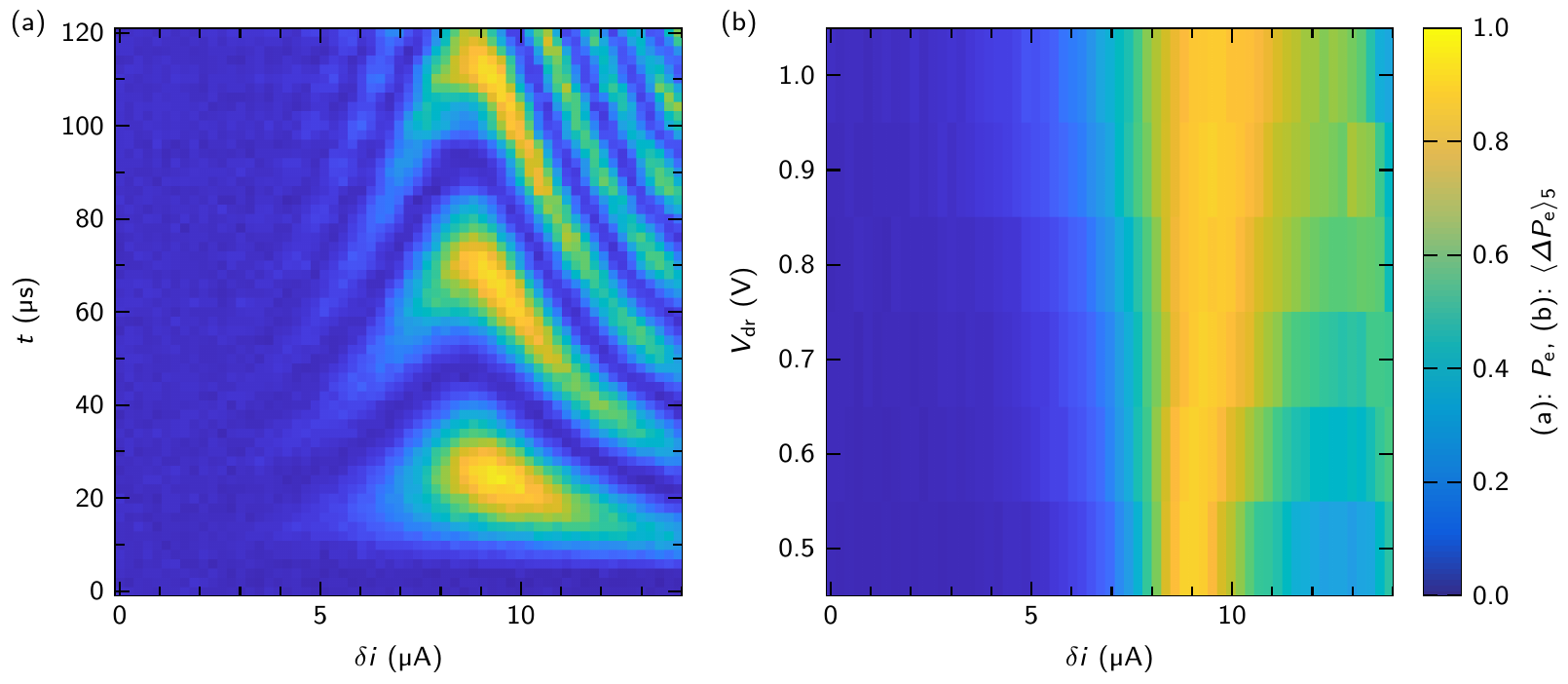}
	\caption{Rotation amplitude control with the DemuXYZ technique. (a)~Heat map of~$P_{\text{e}}$ vs.~$\delta i$ and $t$ for~$V_{\text{dr}}=\SI{0.7}{\volt}$, where~$t$ is the flux-pulse duration. (b)~Heat map of~$\langle \Delta P_{\text{e}} \rangle_5$ vs.~$\delta i$ and $V_{\text{dr}}$. The asymmetry with respect to~$\delta i$ discussed in the main text is also visible in this heat map.}
	\label{Figure05-Bejanin_2022b_Sub}
\end{figure*}

Figure~\ref{Figure03-Bejanin_2022b_Sub}~(b) shows the result of a~PB experiment. Since~$\bar{\mathcal{P}}(m)$ decays faster than~$\bar{P}_{\text{g}}(m)$, we choose~$M=28$ sequence lengths logarithmically spaced between~$1$ and $3000$ Cliffords. The other experimental settings are the same. We fit the~$\bar{\mathcal{P}}$ decay to Eq.~(\ref{eq:PPmbar}) and obtain~$\epsilon_{\text{inc}} = 0.00050(2)$. The total error estimated from the~PB experiment is~$\epsilon = 0.00094(3)$, which is slightly larger than for the~RB experiment of Fig.~\ref{Figure03-Bejanin_2022b_Sub}~(a). These findings lead to~$\epsilon_{\text{coh}} = 0.00044(5)$.

The~RB results indicate that the quantum socket does not impact qubit performance more significantly than standard wirebonds. In particular, the small coherent error rate suggests that unwanted signal reflections (which contribute to part of the coherent error) due to the three-dimensional wires do not pose a major concern for quantum computing applications. This result corroborates the time-domain reflectometry experiments displayed in our previous work of Ref.~\cite{Bejanin:2016}, where we have shown that the wires are characterized by small impedance mismatches. Other sources of coherent errors include frequency mismatch between the microwave drive and the qubit as well as suboptimal pulse shape, which, in principle, could be reduced by improved calibration.

Figure~\ref{Figure04-Bejanin_2022b_Sub}~(a) shows the result of an~RB fluctuation experiment. We choose~$M=19$ sequence lengths logarithmically spaced between~$1$ and $3000$. The execution of a single sequence for each length, that is, one iteration, takes~$\Delta t = \SI{30}{\second}$. Each point~$\bar{F}$ in the figure is the result of a fit performed with a moving average over a~$15$-\si{\minute} window, where each~$\bar{P}_{\text{g}}(m_i)$ value in the~RB exponential decay is thus averaged over~$n=30$ iterations. We measure a total of~$4800$ iterations for an experimental time of~\SI{40}{\hour}. Notably, $\bar{F} > \SI{99.9}{\percent}$ over this entire period. Considering the quantum socket is based on spring-loaded wires and that it is attached to a~DR experiencing mechanical vibrations due to a pulse-tube cryocooler, it is conceivable that the pins could shift with respect to the pads. Pin-pad shifts could negatively impact gate performance due to a changing reflection coefficient. The fluctuation experiment suggests that such mechanical shifts do not pose a concern.

Figure~\ref{Figure04-Bejanin_2022b_Sub}~(b) displays the Allan deviation~(AD)~\cite{Riley:2008} associated with the time series of Fig.~\ref{Figure04-Bejanin_2022b_Sub}~(a). The~AD decreases almost monotonically over the entire analysis interval~$\tau$, indicating near white noise, with a slight flattening at~$\tau \sim \SI{1e4}{\second}$. Overall, there is no clear signature of a~$1/f$-noise behavior, which is often observed in long-time measurements of~$T_1$ and $f_{\text{q}}$. This is likely because the~RB protocol combines a variety of error processes (e.g., energy relaxation, dephasing, and unitary control errors) and twirls the qubit such that its state is not always maximally sensitive to all these errors (e.g., when the qubit is in~$\sim \ket{\text{g}}$, it is mostly insensitive to energy relaxation). Thus, over long time periods, we do not expect the~AD to necessarily follow a~$1/f$ behavior.

\section{\label{sec:DemuXYZ-Based:Gates}DemuXYZ-Based Gates}

\begin{figure*}[!t]
	\centering
	\includegraphics{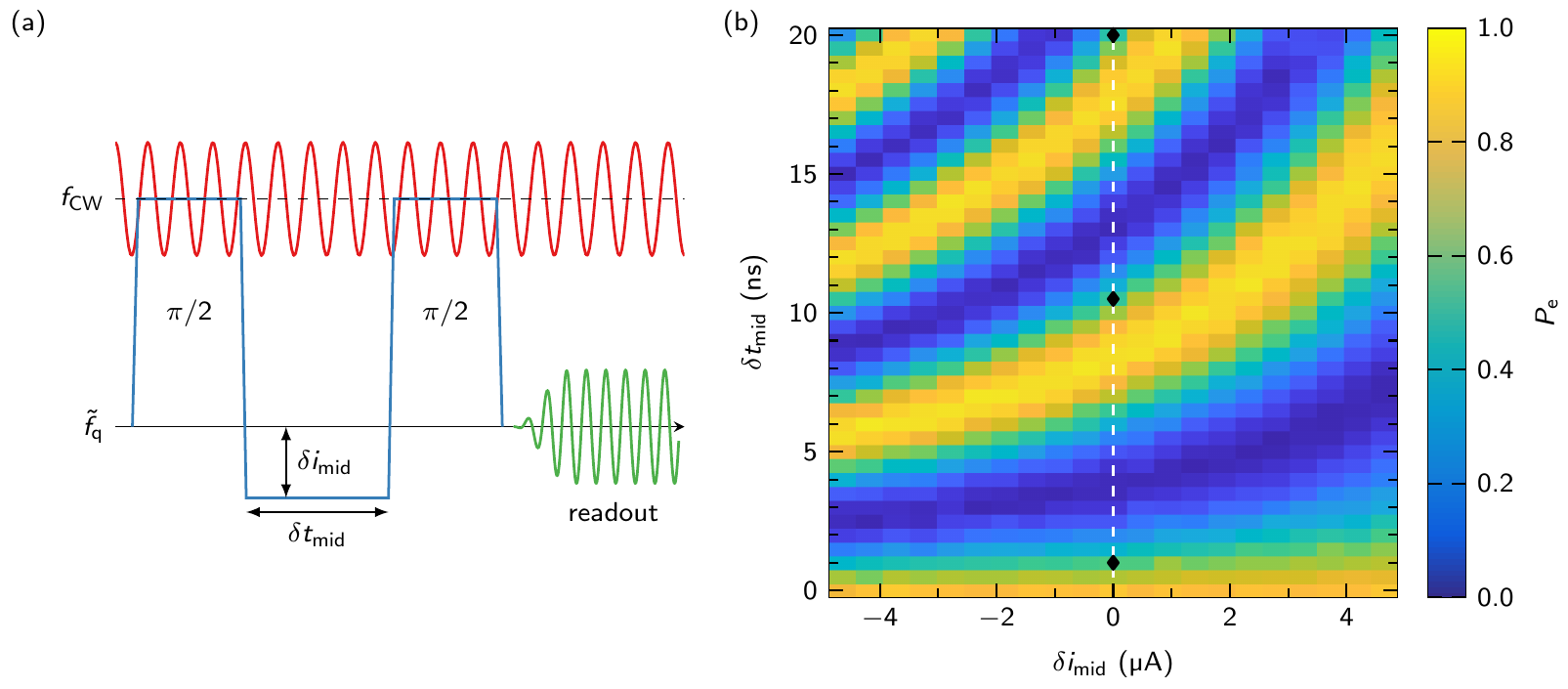}
	\caption{Rotation axis angle control with the DemuXYZ technique. (a)~Ramsey experiment pulse sequence. The rectangular pulses are flux pulses and indicate the trajectory of the instantaneous qubit frequency. We vary the current amplitude and time length of the middle flux pulse, $\delta i_{\text{mid}}$ and $\delta t_{\text{mid}}$. (b)~Heat map of~$P_{\text{e}}$ vs.~$\delta i_{\text{mid}}$ and $\delta t_{\text{mid}}$. The black diamonds along the vertical dashed white line at~$\delta i_{\text{mid}}=\SI{0}{\micro\ampere}$ indicate the points where the relative axis angle of the second~$\pi/2$-pulse is~\SI{90}{\degree}. In this experiment, $\tilde{f}_{\text{q}}=\SI{5.042}{\giga\hertz}$ and $f_{\text{CW}}=\SI{5.147}{\giga\hertz}$; therefore, at~$\delta i_{\text{mid}}=\SI{0}{\micro\ampere}$ the oscillation period is~$1/(f_{\text{CW}}-\tilde{f}_{\text{q}}) \approx \SI{9.5}{\nano\second}$.}
	\label{Figure06-Bejanin_2022b_Sub}
\end{figure*}

\begin{figure}[!hb]
	\centering
	\includegraphics{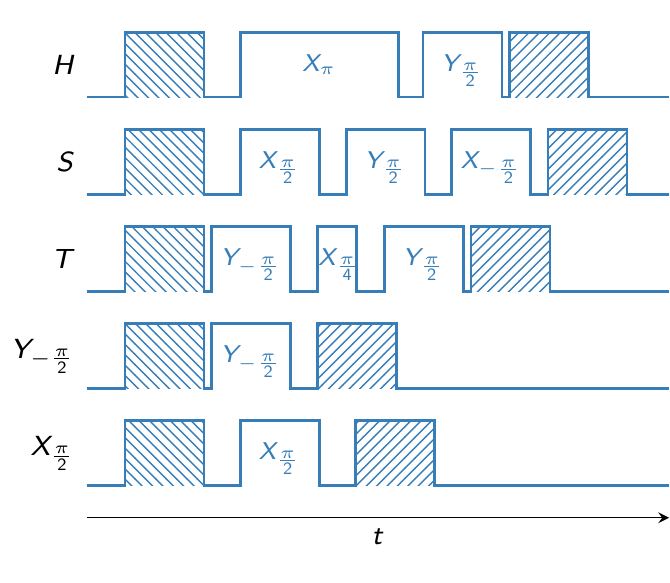}
	\caption{Flux-pulse sequences used to perform~QPT on the DemuXYZ-based gates under test. The first and last pulses are standard microwave pulses used to prepare the initial qubit state and set the measurement basis. The center pulses are DemuXYZ flux pulses (up to three for the~$T$ and $S$ gates) with rectangular shape. These pulses have an identical height chosen to bring the qubit on resonance with~$f_{\text{CW}}$. The pulses' length determines the qubit rotation amplitude and their timing the rotation axis angle. The length of a~$\pi$-pulse is~\SI{40.7}{\nano\second} and the period for a~\SI{360}{\degree} rotation of the axis angle is~\SI{9.9}{\nano\second} (see App.~\ref{app:Calibration:of:DemuXYZ-Based:Gates} for calibration details).}
	\label{Figure07-Bejanin_2022b_Sub}
\end{figure}

DemuXYZ is a control technique engineered to execute~$X$, $Y$, and $Z$ gates on qubits in a demultiplexed fashion, with a single control line simultaneously coupled to multiple qubits. When coupling many qubits to one line, it is not advisable to use microwave pulses for the execution of~$X$ or $Y$ gates. This is because the wide frequency bandwidth of such pulses is likely to affect not only the target qubit but also qubits at nearby frequencies. Bandpass filters could be inserted between the control line and each qubit; however, this solution would necessarily restrict the frequency tunability range of the qubits. In order to demultiplex a control line while preserving qubit tunability, we replace the~$X$ and $Y$ control pulses with a continuous wave~(CW) signal at a fixed frequency~$f_{\text{CW}}$ and flux-pulse the qubit in and out of resonance with that signal.

The DemuXYZ technique can also be used to realize two-qubit gates. A possible protocol would be as follows: A pair of adjacent capacitively coupled qubits is initially prepared with the qubits far off resonance from each other and a common continuous drive. The two qubits are then biased at a given detuning (on the order of~\SI{100}{\mega\hertz}) from each other, while still being detuned from the common drive. Finally, one of the two qubits is brought on resonance with the common drive to implement a cross-resonance gate. A similar gate as been proposed in the work of Ref.~\cite{Allen:2017} using a bus resonator, though it is known that the bus resonator can be replaced by a capacitor~\cite{Patterson:2019}. In this paper, we focus on the experimental implementation of one-qubit DemuXYZ-based gates.

Figure~\ref{Figure05-Bejanin_2022b_Sub} demonstrates control of the qubit rotation amplitude using the DemuXYZ technique. Figure~\ref{Figure05-Bejanin_2022b_Sub}~(a) shows the chevron pattern associated with a qubit Rabi oscillation as obtained with the DemuXYZ technique rather than with standard pulses. At a given qubit detuning (corresponding to a certain~$\delta i$), we control the number of Rabi cycles by setting the duration of the flux pulse. We notice that the pattern is asymmetric with respect to the flux-pulse amplitude corresponding to the middle-point of the chevron (for~$\delta i \approx \SI{9}{\micro\ampere}$). This is likely because the qubit interacts with the~CW signal for different detunings and times during the flux-pulse ramp, whose rise time is set to~\SI{2}{\nano\second} in this experiment (i.e., the flux pulse is not a perfect step function). When pulsing the qubit to a frequency~$f_{\text{q}} > f_{\text{CW}}$, the qubit traverses~$f_{\text{CW}}$ twice; on the other hand, no crossing occurs when the qubit is pulsed to~$f_{\text{q}} < f_{\text{CW}}$. Fitting the Rabi oscillations makes it possible to calibrate the flux-pulse amplitude and length necessary to implement arbitrary rotation amplitudes, such as a~$\pi$-pulse or a~$\pi/2$-pulse.

We then consider each column in the plot of Fig.~\ref{Figure05-Bejanin_2022b_Sub}~(a), take the difference between the five highest and lowest~$P_{\text{e}}$ values for every column, $\langle \Delta P_{\text{e}} \rangle_5$, repeat the experiment at various~CW driving voltage amplitudes~$V_{\text{dr}}$, and plot the results as a heat map in Fig.~\ref{Figure05-Bejanin_2022b_Sub}~(b). This experiment allows us to estimate the on-off ratio of the DemuXYZ technique as the ratio between the highest and lowest values on this heat map; we obtain an on-off ratio of approximately~$45$. It is worth noting that this value is an approximate lower bound since it is largely limited by our readout visibility, which is about~\SI{90}{\percent} for this experiment.

In order to implement arbitrary quantum gates, it is necessary to control the axis about which the qubit state rotates in the Bloch sphere. Standard pulsing techniques allow us to control the axis by changing the relative phase of the drive signal between different pulses. This phase adjustment cannot be performed with the~CW signal used in the DemuXYZ technique. Instead, the rotation axis angle in the~$x$-$y$ plane of the Bloch sphere is controlled by means of a relative time delay between the qubit's reference frame and the leading edge of the flux pulse. Rotations about the~$z$ axis can be realized by simply tuning the qubit away from~$f_{\text{CW}}$.

Figure~\ref{Figure06-Bejanin_2022b_Sub} describes how to control the rotation axis angle using the DemuXYZ technique. Figure~\ref{Figure06-Bejanin_2022b_Sub}~(a) shows the pulse sequence required to implement a DemuXYZ Ramsey experiment. This sequence comprises three flux-pulse steps: The first and last step tune the qubit on resonance with the~CW signal realizing a pair of~$\pi/2$-pulses. The middle step acts as a standard~$Z$ gate during which the qubit remains largely detuned from~$f_{\text{CW}}$ at all times. During the Ramsey experiment, we vary the amplitude and duration~$\delta t_{\text{mid}}$ of the middle step, thereby controlling the relative rotation axis angle of the second~$\pi/2$-pulse.

Figure~\ref{Figure06-Bejanin_2022b_Sub}~(b) shows the results of a DemuXYZ Ramsey experiment. Fitting the Ramsey oscillations allows us to calibrate the time delay corresponding to a particular rotation axis angle. It is important to note that this approach discretizes the time delays available to achieve specific angles. For example, a relative axis angle of~\SI{90}{\degree} (corresponding to an~$X_{\pi/2}$ followed by a~$Y_{\pi/2}$ gate) when~$\delta i_{\text{mid}}=\SI{0}{\micro\ampere}$ is obtained only for time delays~$\delta t_{\text{mid}} \in \{ 1, 10.5, 20,... \} \si{\nano\second}$.

After having explained the working principle of the DemuXYZ technique, we now characterize it quantitatively by means of~QPT~\cite{Nielsen:2000}. Since this is the first implementation of DemuXYZ-based gates, it is important to perform~QPT to ensure the gates properly realize the desired unitary operations.

A DemuXYZ-based gate comprises one or multiple flux pulses. To realize and optimize DemuXYZ-based gates a careful calibration is required. For each pulse three parameters must be determined: (1)~The flux-pulse amplitude. (2)~The flux-pulse duration. (3)~The timing of the leading edge of the flux pulse. The calibration details are provided within App.~\ref{app:Calibration:of:DemuXYZ-Based:Gates}.

We implement~QPT by preparing the initial qubit state, applying the desired gate, and, finally, setting the measurement basis before readout; all the operations are carried out using DemuXYZ flux pulses. We use the six axial Bloch sphere states as input states and then readout in the~$\mp \hat{\sigma}_{x,y,z}$ bases for a total of~$36$ measurements per gate under test. We test five gates, $X_{\pi/2}, Y_{-\pi/2}, T, S$, and $H$, and then reconstruct the Choi process matrix~$\mathcal{C}$ using a projected gradient descent algorithm~\cite{Knee:2018}.

Figure~\ref{Figure07-Bejanin_2022b_Sub} displays the flux-pulse sequences used to realize the five gates under test. Simple gates such as~$X_{\pi/2}$ and $Y_{\pi/2}$ necessitate only a single flux pulse, where the pulse duration determines the rotation axis angle. While it is possible to implement~$T$ and $S$ gates with rotations about the~$z$ axis by simply tuning the qubit away from~$f_{\text{CW}}$, we choose to decompose the~$T$ and $S$ gates only in terms of resonant DemuXYZ $X$ (or $Y$) gates and not to use~$Z$ gates to implement the~$H$ gate. This allows us to test multi-pulse DemuXYZ-based gates on resonance with~$f_{\text{CW}}$, rather than well-known~$Z$ gates.

The average gate fidelity~$\bar{F}$ for the five tested DemuXYZ-based gates is reported in Table~\ref{tab:QPT:Fidelity}. The fidelities are obtained from the reconstructed~$\mathcal{C}$ matrices, which are reported in Sec.~S2 within the Supplemental Material~\cite{SM}.

\begin{table}[!t]
 \caption{Average gate fidelity~$\bar{F}$ obtained from~$\mathcal{C}$ for the five tested DemuXYZ-based gates. As a comparison, we also report the average fidelity for gates realized by means of standard microwave pulses. \label{tab:QPT:Fidelity}}
   \begin{ruledtabular}
    \begin{tabular}{lccccc}
     \raisebox{0mm}[2mm][0mm]{Gate} & $X_{\pi/2}$ & $Y_{-\pi/2}$ & $T$ & $S$ & $H$ \\
     \hline
     \raisebox{0mm}[3mm][0mm]{$\bar{F}$ DemuXYZ} & \SI{95.65}{\percent} & \SI{96.23}{\percent} & \SI{93.75}{\percent} &\SI{88.93}{\percent} & \SI{91.36}{\percent} \\
     \hline
     \raisebox{0mm}[3mm][0mm]{$\bar{F}$ standard} & \SI{98.67}{\percent} & \SI{98.18}{\percent} & \SI{98.12}{\percent} &\SI{98.39}{\percent} & \SI{98.66}{\percent} \\
    \end{tabular}
   \end{ruledtabular}
\end{table}

\section{\label{sec:Discussion}Discussion}

In Sec.~\ref{subsec:Scalable:wiring}, we compare the quantum socket with state-of-the-art wiring techniques and propose future improvements. In Sec.~\ref{subsec:Multiplexing}, we address open problems related to qubit control demultiplexing.

\subsection{\label{subsec:Scalable:wiring}Scalable wiring}

In recent years, several extensible interconnects have been developed to connect a qubit chip to an external control and measurement network. In the work of Ref.~\cite{Bronn:2018}, spring-loaded Pogo pins are used to connect a signal board to a chip with superconducting transmon qubits. The pins reach the qubit chip through holes within a copper block acting as an interposer, thereby shielding the qubits from the signal board. While this approach is based on spring-loaded wires similar to ours (although not coaxial), its scalability is highly limited by the signal board that requires lateral wire bonds to connect to the external network.

Interposers have since been engineered to a higher level of integrability, for example, in the work of Ref.~\cite{Yost:2020}. In that work, the Pogo pins are replaced by metallized superconducting through-silicon vias; the shielding is now provided by metallizing one side of the silicon interposer chip. This technique allows for three types of electrical couplings to the qubit chip: capacitive, inductive, and galvanic. Capacitive and inductive couplings are realized across the vacuum gap between the qubit chip and the interposer, whereas galvanic couplings can be achieved by means of superconducting indium bump bonds~\cite{Rosenberg:2017}. It is worth noting that several groups have elected not to use an interposer and, instead, directly bond together the qubit and signal chips using flip-chip techniques~\cite{Foxen:2017,Kosen:2022}. All these methods, however, still rely on lateral wire bonds to connect to the external network and, thus, are likely not scalable beyond a few hundred qubits.

A different wiring technique based on coaxial cables hanging directly above a qubit chip has been demonstrated in the works of Refs.~\cite{Rahamim:2017,Spring:2022}. Like the quantum socket, this interconnect is truly vertical and, thus, is potentially more scalable than current flip-chip techniques. Although the coaxial-cable method does not allow for galvanic couplings, it can be used for both capacitive and inductive couplings, which are sufficient to run a superconducting quantum computer; however, the physical dimensions of the cables, on the millimeter range, are far too large to scale to more than a few tens of qubits. Indeed, the present implementation of the quantum socket suffers from a similar limitation.

Although too large for scalability, the quantum socket and coaxial-cable techniques are characterized by a critical common feature, that is, the physical footprint of the external network is similar to that of the interconnect itself; this is due to their vertical implementation. A truly scalable wiring architecture requires both small (on the order on tens of micrometers) and vertical interconnects, as well as equally small signal lines along the external network.

As an integrated and miniaturized implementation of the quantum socket, we propose a truly scalable wiring architecture based on stripline transmission lines microfabricated on flexible polyimide ribbons~\cite{Tuckerman:2016}. In this case, the interconnect can be realized as an extension of the ribbon cables, thus leading to an identical footprint of the interconnect and external network. In one approach, we would terminate the striplines in open or short circuits at a suitable distance from the qubit chip, leading to capacitive or inductive couplings. The coupling mechanisms of this approach are conceptually the same as in the work of Ref.~\cite{Spring:2022}; however, the overall wiring solution would be significantly denser and more integrated. An alternative approach allowing for galvanic (superconducting) couplings similar to the quantum socket would rely on terminating the ribbon cables into micromachined rigid pins. Instead of using springs to provide a suitable force at the pin-pad connection, the pins would pierce into an array of indium bumps fabricated on the qubit chip~\cite{Mariantoni:2018}. Both approaches can be miniaturized such that each signal line has a similar footprint of a physical qubit (i.e., approximately~\SI{100}{\micro\meter}) all the way from the qubit chip to room temperature.

It is worth noting that ribbon cables can also be used to directly integrate a variety of control and measurement devices that would otherwise be fabricated on the qubit or signal chips or placed along the external network as bulky components. Such devices comprise readout resonators, attenuators (and, thus, signal-line thermalization), filters, amplifiers, and possibly rapid single flux quantum electronics. This integration would make it possible to significantly miniaturize the footprint of microwave electronics and to efficiently use the vertical space above the qubit chip; it would also reduce the amount of manual work required to assemble the wiring of a dilution refrigerator.

\subsection{\label{subsec:Multiplexing}Multiplexing}

In the realm of superconducting quantum computing, (de)multiplexing techniques are still in their infancy. Among the few experimental implementations, we point out the ubiquitous frequency-multiplexed readout of qubits~\cite{Kjaergaard:2020} and the operation of a quantum computer with up to four qubits residing within a single bus resonator~\cite{Reed:2012}. A two-dimensional grid of such bus resonators has been proposed to implement a scaled up version of that work~\cite{Helmer:2009:a}. Finally, spatial demultiplexing based on a fast switching matrix has been recently investigated in the work of Ref.~\cite{Versluis:2017}.

Ideally, the DemuXYZ technique makes it possible to reduce the number of wires for frequency-tunable superconducting qubits from~$2N$ [$N$ wires for~$Z$ gates and $N$ wires for~$X$ (or~$Y$) gates] to $N+1$ ($N$ wires for~$Z$ gates and one wire for the common signal drive). This technique can easily be adapted to different qubit layouts, for example, with blocks of qubits residing along a single meandered transmission line or grouped around a central node. Additionally, multiple simultaneous~CW signals could be employed to further scale up the DemuXYZ technique.

It is worth noting that a similar demultiplexing factor as that obtained using the DemuXYZ technique can be achieved by combining standard~$X$ and $Y$ control lines with~$Z$ lines into a single line. This approach would require the careful engineering of integrated filters to accurately separate microwave from flux pulses (i.e., a diplexer) at the qubit level. Indeed, such filters could be integrated within the ribbon cables proposed in the previous section. Nevertheless, we think that a higher level of demultiplexing is eventually required to reach true scalability. One possibility to demultiplex microwave pulses is to use a two-dimensional grid similar to that in the work of Ref.~\cite{Helmer:2009:a}, where, however, a combination of two distinct pulses is needed in order to address a single qubit without interacting with any other qubit. It would be important to investigate such an approach as it could lead to a much larger demultiplexing factor, from~$N^2$ wires to just~$2N$ wires for~$N^2$ qubits. Similarly, a two-dimensional grid could be used to significantly reduce the number of flux lines required to implement DemuXYZ-based gates.

\section{\label{sec:Conclusions}Conclusions}

In this paper, we first investigate the quantum socket using a long-coherence Xmon transmon qubit. We test the near-dc performance of the socket by tuning the qubit frequency with both a quasistatic flux bias and flux pulses. We do not detect any heating generated by the pin-pad resistance of the three-dimensional wires in the quantum socket, even at the highest current of approximately~\SI{100}{\micro\ampere}. This finding indicates that a galvanic pin-pad connection without cold welding, as with wire bonds or flip-chip techniques, is suitable to bias and control superconducting qubits. We test gate performance by means of~RB and PB and show that one-qubit gates with high fidelity are achievable, even when using short microwave pulses (\SI{20}{\nano\second}) without buffer time, hinting at negligible signal reflections. We additionally confirm the time stability of one-qubit gates realized with the quantum socket by measuring an~RB fidelity in excess of~\SI{99.9}{\percent} for~\SI{40}{\hour}. All these results indicate that our three-dimensional wiring approach is viable for scaling up toward large-scale quantum computers, given that further miniaturization and integration is possible with present technology.

We then introduce a demultiplexed technique, DemuXYZ, to implement one- and two-qubit gates. We explain how the rotation axis amplitude and angle are controlled. We finally characterize a selection of DemuXYZ-based gates by means of one-qubit~QPT, obtaining fidelities between~$89$ and \SI{96}{\percent}. The fidelity of a DemuXYZ-based gate strongly depends on the accuracy of the flux~$Z$ pulses necessary to implement it. High-accuracy flux pulses can be achieved by engineering them both at the qubit level, as proposed in the works of Refs.~\cite{Galiautdinov:2012,Egger:2013}, and accounting for the imperfections of the external network leading to the qubits as, for example, explored in the works of Refs.~\cite{Spoerl:2007,Bylander:2009}. Future work will focus on implementing such improvements as well as testing DemuXYZ-based gates in multi-qubit systems.

\begin{table*}[ht!]
	\centering
	\caption{Exemplary decomposition for each of the~$24$ one-qubit Clifford gates. The notation~$R_{\vec{u}}(\theta)$ represents a Clifford gate as a Bloch-sphere rotation by an angle~$\theta$ about the axis represented by the vector~$\vec{u}=(x,y,z)$ in a Cartesian coordinate system. The right column presents a decomposition of the Clifford gate in terms of primitive gates. Gates in the decomposition are applied in the listed order (left to right).}
	\label{tab:Clifford:Gate:Decomposition}
	\begin{minipage}{0.8\linewidth}
		\begin{ruledtabular}
			\begin{tabular}[t]{llll}
				Clifford gate & Decomposition \hspace{1cm} & Clifford gate & Decomposition \\ \hline \\[-3mm]
				$R_{(1,0,0)}(0)$         & $I$              & $R_{(1,0,0)}(\pi/2)$     & $X_{\pi/2}$ \\
				$R_{(1,0,0)}(\pi)$       & $X_{\pi}$              & $R_{(-1,0,0)}(\pi/2)$    & ${X_{-\pi/2}}$ \\
				$R_{(0,1,0)}(\pi)$       & $X_{\pi},~Z_{\pi}$           & $R_{(0,1,0)}(\pi/2)$     & $X_{\pi/2},~Z_{\pi/2},~{X_{-\pi/2}}$ \\
				$R_{(0,0,1)}(\pi)$       & $Z_{\pi}$              & $R_{(0,-1,0)}(\pi/2)$    & $X_{\pi/2},~{Z_{-\pi/2}},~{X_{-\pi/2}}$ \\
				$R_{(1,1,1)}(2\pi/3)$    & $X_{\pi/2},~Z_{\pi/2}$       & $R_{(0,0,1)}(\pi/2)$     & $Z_{\pi/2}$ \\
				$R_{(1,1,-1)}(2\pi/3)$   & ${Z_{-\pi/2}},~X_{\pi/2}$    & $R_{(0,0,-1)}(\pi/2)$    & ${Z_{-\pi/2}}$ \\
				$R_{(1,-1,1)}(2\pi/3)$   & $Z_{\pi/2},~X_{\pi/2}$       & $R_{(1,0,1)}(\pi)$       & $X_{\pi/2},~Z_{\pi/2},~X_{\pi/2}$ \\
				$R_{(1,-1,-1)}(2\pi/3)$  & $X_{\pi/2},~{Z_{-\pi/2}}$    & $R_{(1,0,-1)}(\pi)$      & $X_{\pi/2},~{Z_{-\pi/2}},~X_{\pi/2}$ \\
				$R_{(-1,1,1)}(2\pi/3)$   & $Z_{\pi/2},~{X_{-\pi/2}}$    & $R_{(0,1,-1)}(\pi)$      & $Z_{\pi},~X_{\pi/2}$ \\
				$R_{(-1,1,-1)}(2\pi/3)$  & ${X_{-\pi/2}},~{Z_{-\pi/2}}$ & $R_{(0,1,-1)}(\pi)$      & $Z_{\pi},~{X_{-\pi/2}}$ \\
				$R_{(-1,-1,1)}(2\pi/3)$  & ${X_{-\pi/2}},~Z_{\pi/2}$    & $R_{(1,1,0)}(\pi)$       & $X_{\pi},~Z_{\pi/2}$ \\
				$R_{(-1,-1,-1)}(2\pi/3)$ & ${Z_{-\pi/2}},~{X_{-\pi/2}}$ & $R_{(-1,1,0)}(\pi)$      & $X_{\pi},~{Z_{-\pi/2}}$ \\
			\end{tabular}
		\end{ruledtabular}
	\end{minipage}
\end{table*}

\begin{figure*}[!t]
	\centering
	\includegraphics{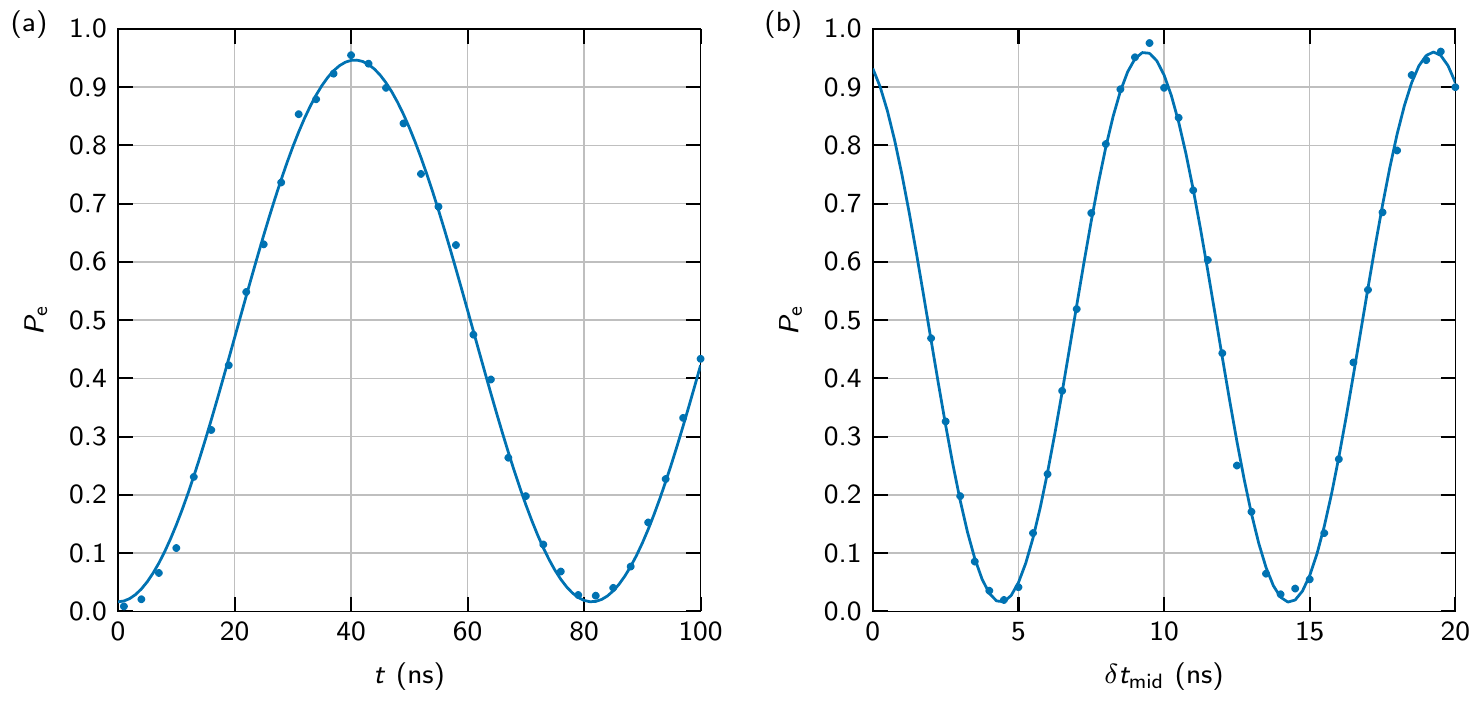}
	\caption{Calibration of DemuXYZ-based gates for~QPT experiments. (a)~$P_{\text{e}}$ vs.~$t$. The solid line is a sinusoidal fitting curve with a Rabi frequency of~\SI{12.30(2)}{\mega\hertz}. (b)~$P_{\text{e}}$ vs.~$\delta t_{\text{mid}}$. The solid line is a cosine fit with a Ramsey frequency of~\SI{101.0(2)}{\mega\hertz} and a phase of~\SI{0.35(2)}{\radian}.
		\label{Figure08-Bejanin_2022b_Sub}}
\end{figure*}

\begin{acknowledgments}
This research was undertaken thanks in part to funding from the Canada First Research Excellence Fund~(CFREF). We acknowledge the support of the Natural Sciences and Engineering Research Council of Canada~(NSERC), [Application Number: RGPIN-2019-04022]. We would like to acknowledge CMC Microsystems and Canada's National Design Network~(CNDN). The authors thank the Quantum-Nano Fabrication and Characterization Facility at the University of Waterloo, where the samples were fabricated.
\end{acknowledgments}

\appendix

\section{\label{app:Clifford:Gate:Decomposition}Clifford Gate Decomposition}

The two main types of one-qubit physical gates available with superconducting qubits are: (1) Microwave-based~$X$ (or $Y$) gates, which directly rotate the qubit state about any axis in the $x$-$y$~plane of the Bloch sphere ($x$-$y$ rotations). (2) Flux-based~$Z$ gates, which tune~$f_{\text{q}}$, thereby leading to a rotation about the~$z$ axis when referenced to~$\tilde{f}_{\text{q}}$. Arbitrary one-qubit gates, however, may require rotations about any (e.g., tilted) axes in the Bloch sphere. For example, the Hadamard gate is a~$\pi$ rotation about the axis~$\vec{u}=(x=1,y=0,z=1)$. Consequently, the Hadamard gate or any other gates needing tilted rotation axes cannot be implemented as a single physical gate; instead, they need to be decomposed into multiple physical gates.

In order to decompose a given Clifford gate, we must therefore choose a set of primitive physical gates. For example, with an all-microwave gate set~$\{I,X_{\pi},X_{\pi/2},Y_{\pi},Y_{\pi/2}\}$, where~$I$ is the identity gate, we may decompose the Hadamard gate into two primitive gates as~$H=i \, X_{\pi} \, Y_{\pi/2}$. If, instead, we choose to use flux-based~$Z$ gates, a possible decomposition is~$H=i \, X_{\pi/2} \, Z_{\pi/2} \, X_{\pi/2}$, that is, a flux pulse in between two microwave pulses. Yet another approach replaces physical~$Z$ gates with \emph{virtual}~$Z$~gates. Rotations about the~$z$ axis are equivalent to phase shifts of the qubit's reference frame; thus, it is possible to execute virtual~$Z$ gates by tracking the reference frame's phase and rotating the axis of subsequent~$x$-$y$ rotations~\cite{McKay:2017}. In this case, the Hadamard gate is once again implemented with two microwave pulses, where the axis of the second pulse is itself rotated by~\SI{90}{\degree}.

In our experiments, we choose a primitive gate set based on microwave~$X$ and virtual~$Z$ gates. The set comprises the following seven primitive gates: $\{ I , X_{\pi} , X_{\mp \pi/2} , Z_{\pi} , Z_{\mp \pi/2} \}$. This gate set does not lead to a unique decomposition for each Clifford gate. Table~\ref{tab:Clifford:Gate:Decomposition} presents an exemplary decomposition for each of the~$24$ one-qubit Clifford gates.

\section{\label{app:Calibration:of:DemuXYZ-Based:Gates}Calibration of DemuXYZ-Based Gates}

The first step [step~(1) in the main text] in the calibration of a DemuXYZ-based gate is to determine the proper flux-pulse amplitude to bring the qubit on resonance with the signal drive at~$f_{\text{CW}}$ starting from~$\tilde{f}_{\text{q}}$. The experimental procedure is the same as that used to find the results in Fig.~\ref{Figure05-Bejanin_2022b_Sub}, that is, we measure multiple Rabi oscillations for different flux amplitudes (i.e., $\delta i$) and fit them to determine the point where the oscillation amplitude is maximized.

In order to calibrate the flux-pulse duration [step~(2)], we run another Rabi experiment at the flux-pulse amplitude determined in step~(1). Once again, we obtain the duration by fitting the Rabi oscillation, as shown in Fig.~\ref{Figure08-Bejanin_2022b_Sub}~(a).

Finally, we must calibrate the timing of the leading edge of each flux pulse [step~(3)]. The experimental procedure is the same as that used to find the results in Fig.~\ref{Figure06-Bejanin_2022b_Sub} for~$\delta i_{\text{mid}}=0$ (i.e., we return to~$\tilde{f}_{\text{q}}$ in between the pulses). We then fit the data, as shown in Fig.~\ref{Figure08-Bejanin_2022b_Sub}~(b). When~$P_{\text{e}}=1$, the rotation axis is the same for both pulses, when~$P_{\text{e}}=0$ the axes have opposite directions, and when~$P_{\text{e}}=1/2$ they are normal to each other. When setting the rotation axis angle of each flux pulse in a DemuXYZ sequence (i.e., the timing of the leading edge), we must account for the discrete nature of the time delays available to achieve specific angles. Since the time delays are periodic, for each pulse in the sequence we choose the first available delay that does not result in an overlap with the previous pulse in the sequence.

\bibliography{bibliography}

\clearpage

\pagebreak

\begin{center}
	\textbf{\large Supplemental Material for ``The Quantum Socket and DemuXYZ-Based Gates with Superconducting Qubits''}
\end{center}

\newcommand{\beginsupplement}{%
\setcounter{section}{0}
\renewcommand{\thesection}{S\arabic{section}}%
\setcounter{subsection}{0}
\renewcommand{\thesubsection}{S\Roman{subsection}}%
\setcounter{subsubsection}{0}
\renewcommand{\thesubsubsection}{S\Alph{subsubsection}}%
\titleformat{\subsubsection}[block]{\bfseries\centering}{\thesubsubsection.}{1em}{}
\setcounter{table}{0}
\renewcommand{\thetable}{S\arabic{table}}%
\setcounter{figure}{0}
\renewcommand{\thefigure}{S\arabic{figure}}%
\setcounter{equation}{0}
\renewcommand{\theequation}{S\arabic{equation}}%
}

\begin{widetext}
\section*{S1: Qubit Parameters}

The Xmon transmon qubits measured in this work are fabricated using the same procedure as explained in our previous works of Refs.~\cite{Bejanin:2021:a,Bejanin:2021:b}; the measurement setup is shown in App.~A of Ref.~\cite{Bejanin:2021:b}. The parameters of the qubit used for the heating and DemuXYZ experiment are reported in App.~A of Ref.~\cite{Bejanin:2021:b}. For the RB~experiments we use a different qubit with~$\tilde{f}_{\text{q}} \approx \SI{4.8366}{\giga\hertz}$ and a readout resonator with the resonance frequency of~\SI{5.2988}{\giga\hertz}; additionally, the measurement visibility for this qubit is approximately~\SI{90}{\percent}.

Figure~\ref{Figure01_SM-Bejanin_2022b_Sub} shows standard qubit performance measurements, namely of qubit energy relaxation~$T_1$, qubit dephasing~$T_2$, and qubit dephasing with echo sequence, $T_2^{\text{echo}}$, for the qubit used in the RB experiment. These experiments are performed immediately after the~$40$-\si{\hour} RB temporal stability experiment and, therefore, are representative of the qubit performance during~RB (considering the~RB fidelity is very stable throughout the entire experiment).

\begin{figure*}[ht!]
	\centering
	\includegraphics{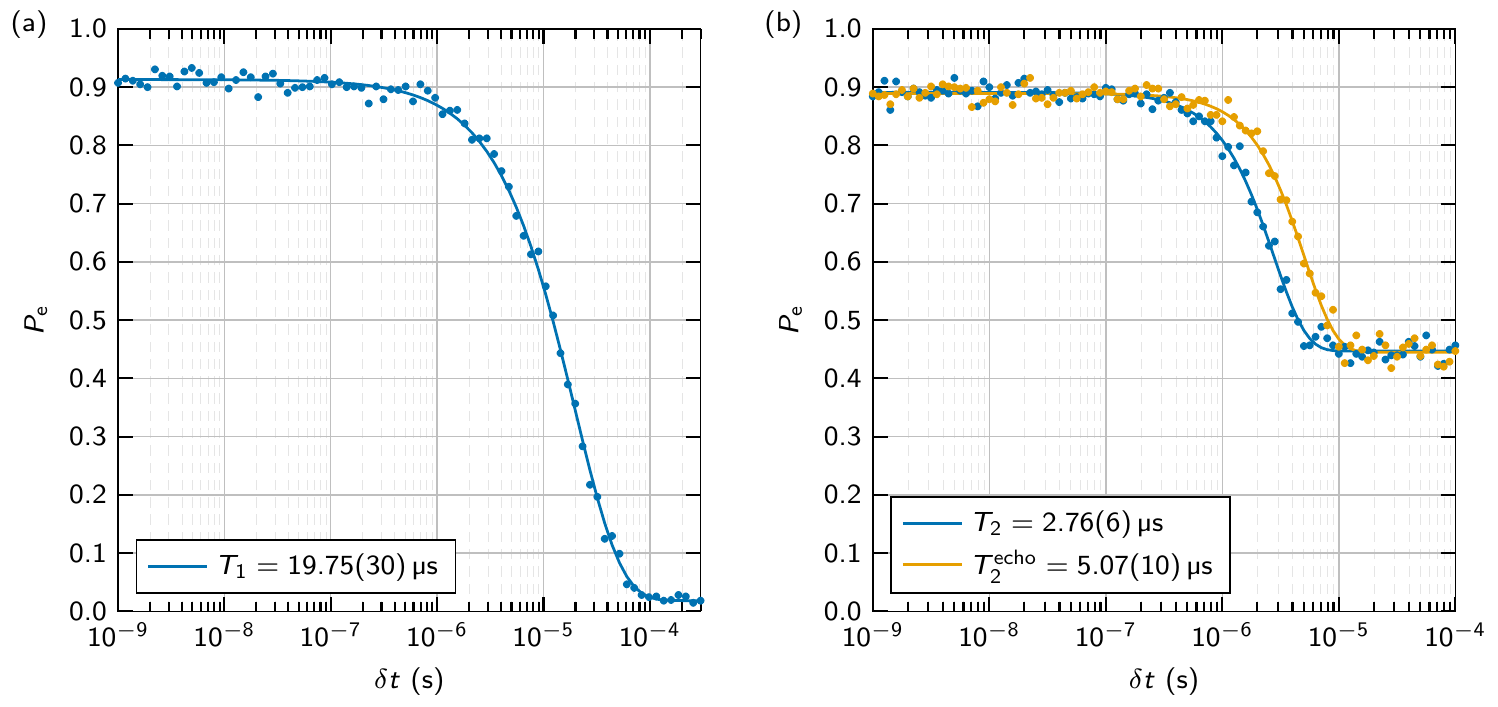}
	\caption{Standard qubit performance measurements. (a),(b)~$P_{\text{e}}$ vs.~$\delta t$. Dots correspond to experimental values and solid lines to fitting curves from a simple exponential decay model. The fitting parameters are displayed within the legends. The large improvement in the dephasing time with the echo sequence indicates that this qubit is affected by a significant amount of low-frequency noise.
		\label{Figure01_SM-Bejanin_2022b_Sub}}
\end{figure*}

\section*{S2: Choi Process Matrices}

The reconstructed Choi process matrices are displayed in the following equations. For the~$X_{\pi/2}$ gate, we obtain the~$\mathcal{C}_{X_{\pi/2}}$ matrix,
\begin{equation}
	\begin{pmatrix}
		0.512912              &&   0.138107+0.437984i     &&  -0.0608019+0.458569i   &&   0.484966-0.00862546i \\
		0.138107-0.437984i    &&   0.487088               &&   0.43032+0.166958i     &&   0.0608019-0.458569i \\
		-0.0608019-0.458569i   &&   0.43032-0.166958i      &&   0.46196               &&  -0.109849-0.468808i \\
		0.484966+0.00862546i  &&   0.0608019+0.458569i    &&  -0.109849+0.468808i    &&   0.53804
	\end{pmatrix} ,
\end{equation}
for the~$Y_{-\pi/2}$ gate, the~$\mathcal{C}_{Y_{-\pi/2}}$ matrix,
\begin{equation}
	\begin{pmatrix}
		0.574053               &&  -0.427494+0.0518717i    &&  0.477174+0.0214284i    &&  0.529006+0.00781682i \\
		-0.427494-0.0518717i    &&   0.425947               && -0.398762-0.0990363i    && -0.477174-0.0214284i \\
		0.477174-0.0214284i    &&  -0.398762+0.0990363i    &&  0.433599               &&  0.464487-0.0596072i \\
		0.529006-0.00781682i   &&  -0.477174+0.0214284i    &&  0.464487+0.0596072i    &&  0.566401
	\end{pmatrix} ,
\end{equation}
for the~$T$ gate, the~$\mathcal{C}_{T}$ matrix,
\begin{equation}
	\begin{pmatrix}
		0.910526              &&   0.120484+0.0271023i    &&  -0.0439363+0.0751614i  &&   0.805124-0.425477i \\
		0.120484-0.0271023i   &&   0.0894741              &&   0.0140933-0.0113732i  &&   0.0439363-0.0751614i \\
		-0.0439363-0.0751614i  &&   0.0140933+0.0113732i   &&   0.0256934             &&  -0.100904-0.0602796i \\
		0.805124+0.425477i    &&   0.0439363+0.0751614i   &&  -0.100904+0.0602796i   &&   0.974307
	\end{pmatrix} ,
\end{equation}
for the~$S$ gate, the~$\mathcal{C}_{S}$ matrix,
\begin{equation}
	\begin{pmatrix}
		0.80507                &&   0.075515-0.20558i      &&  -0.156325+0.0324483i   &&    0.244115-0.801172i \\
		0.075515+0.20558i      &&   0.19493                && -0.0304349-0.0361021i   &&    0.156325-0.0324483i \\
		-0.156325-0.0324483i    &&  -0.0304349+0.0361021i   &&  0.0713923              &&  -0.0979701+0.160685i \\
		0.244115+0.801172i     &&   0.156325+0.0324483i    && -0.0979701-0.160685i    &&    0.928608
	\end{pmatrix} ,
\end{equation}
and, finally, for the~$H$ gate, the~$\mathcal{C}_{H}$ matrix,
\begin{equation}
	\begin{pmatrix}
		0.477031               &&   0.438883+0.0252139i    &&  0.398773+0.19267i      && -0.271784-0.200279i \\
		0.438883-0.0252139i    &&   0.522969               &&  0.504164+0.193168i     && -0.398773-0.19267i \\
		0.398773-0.19267i      &&   0.504164-0.193168i     &&  0.575809               && -0.469063-0.0555586i \\
		-0.271784+0.200279i     &&  -0.398773+0.19267i      && -0.469063+0.0555586i    &&  0.424191
	\end{pmatrix} .
\end{equation}
\end{widetext}


\end{document}